\documentclass[structabstract]{aa}
\usepackage{graphicx}
\usepackage{txfonts}
\usepackage[]{natbib}
\usepackage{longtable}
\usepackage{color}
\usepackage{slashbox}
\usepackage{multirow}
\begin{document}
   \title{The analysis of the hydrogen broad Balmer line ratios: possible\\
 implications to the physical properties of the BLR of AGN}

   \subtitle{}

   \author{D. Ili\'c
          \inst{1,2}
          \and
          L. \v C. Popovi\'c\inst{3,2}
          \and
          G. La Mura\inst{4}
          \and
          S. Ciroi\inst{4}
          \and P. Rafanelli\inst{4}
          }

   \institute{Department of Astronomy, Faculty of Mathematics,
   University of Belgrade, Studentski trg 16, 11000 Belgrade, Serbia
              \email{dilic@matf.bg.ac.rs}
         \and
              Isaac Newton Institute of Chile, Yugoslavia Branch, Belgrade, Serbia
         \and
              Astronomical Observatory, Volgina 7, 11160 Belgrade, Serbia
         \and
              Dipartimento di Fisica e Astronomia, Universit\`a di Padova, 
              Vicolo dell'Osservatorio, I-35122 Padova, Italy
             }

   \date{Received xxxx, 2011; accepted xxxx, xxxx}

 \titlerunning{The analysis of the hydrogen broad Balmer line ratios}


  \abstract
   {}
   {We analyze the ratios of the broad hydrogen Balmer emission lines (from H$\alpha$ to 
H$\varepsilon$) in the context of estimating the physical conditions 
in the broad line region (BLR) of active galactic nuclei (AGN). }
   {The Balmer emission lines are obtained in three ways: i) using photoionization 
models obtained by a spectral synthesis code CLOUDY;  ii) calculated  using the recombination
theory for hydrogenic ions; iii) measured from the sample of observed spectra
taken from the Sloan Digital Sky Survey database. We investigate the Balmer
line ratios in the frame of the so called Boltzmann-plot (BP), analyzing
physical conditions of the emitting plasma for which we could use the BP method.
The BP considers the ratio of Balmer lines normalized to the atomic data of 
the corresponding line transition, and is in that way different from the Balmer decrement.}
   {We found that for a certain range of thermodynamic parameters, there are objects 
that follow the BP. These AGN may have a BLR consisting of mostly high density plasma.}
   {}


   \maketitle
%

\section{Introduction}

The investigations of the broad line region (BLR) structure (size, geometry, 
physics, etc.) of active galactic nuclei (AGN) are nowadays still important. 
One of the main implications is the accurate estimate of the mass of 
the super-massive black hole (SMBH) in the center of an AGN, since it can be 
derived from the dynamics of BLR gas, gravitationally bound to the SMBH  
\citep[see e.g.][]{M08,Be10}. After several decades of research, 
our view of the physics and structure of the BLR is improving, but we still lack a 
full understanding of its properties \citep[see e.g.][for a review]{Ba97, Su00, Ga09}. 
It is obvious from the broad emission line (BEL) fluxes and profiles that the 
BLR physics and kinematics are more complicated than in the narrow line region (NLR) or 
other nebular environments \citep[see e.g.][and references therein]{Kr99, Su00, OF06}, 
and they are even closer to the conditions in stellar atmospheres than in 
gaseous nebulae \citep{OF06}.

Most of the BLR gas is photoionized. This was first inferred by the approximately 
constant value of the H$\beta$ equivalent width in several AGN \citep{Os78} and then 
confirmed by reverberation studies showing a direct response of the BEL fluxes to 
the continuum flux changes \citep[see e.g.][]{Pe94}. 
So far there have been numerous attempts to describe the BLR physics and to explain 
the emission-line spectrum by means of photoionization models \citep[see e.g.][]{Ne75, 
DN79, KK81, CS82, RNF89, Ba95, Ba96, Du98, Fe98, Kr99, Ma01, Fe03, KG04, LC07, SG07, Ma10}.
Today there are roughly half a dozen major codes, that were continuously developed 
since the 1970s \citep[see e.g.][for a review]{FS01}. Although such numerical 
simulations largely improved in the course of years, the investigations of the BLR 
still contain some important unresolved problems \citep[see e.g.][]{Fe99}. One of the most 
interesting concerns the observed small ratio among the Ly$\alpha$ and H$\beta$ lines 
(usually in the range of 5 - 15), while models predict much higher values (30 - 50) 
\citep[see e.g.][and references therein]{Ne95}. Another important finding is
that there are indications for a physical complexity of the BLR, which could 
result from the combination of distinct components, as suggested, for instance, by the 
origin of ``high'' and ``low ionization lines'' \citep[see e.g.][]{CS88, Ba97, Su00}, 
but also according to the complex profiles of single lines, that cannot be explained in terms 
of a simple single-region model \citep[see e.g.][]{Po04, Po06b, Bo09, Ga09}. 

Different BELs are originated at various distances from 
the central continuum source and under a wide range of physical conditions 
\citep[see e.g.][]{Su00}. The BLR is of high density, thus the collisional 
excitation, self-absorption, dust obscuration and complicated coupling of line 
and continuum radiation transfer should be taken into account in the calculation 
of the resulting spectrum \citep[see e.g.][]{Ba97}. 
In such circumstances, the general spectroscopic techniques, commonly adopted to 
estimate physical conditions in nebular environments, cannot be applied. Some clues, 
such as the observed Fe II emission, suggest that $T\leq35,000$ K, since at higher 
temperatures it would be effectively suppressed by collisional ionization to 
Fe III \citep{OF06}. However it is very likely that the Fe II emission only affects a 
fraction of the BLR \citep[see e.g.][and references therein]{Ma01, Il09, Po09, Ko10}. 
The electron density is estimated to be in the range $ 10^8 {\rm cm ^{-3}} \lesssim
n_{\rm e}  \lesssim 10^{14} {\rm cm ^{-3}}$, in order to suppress the emission 
of broad forbidden lines, while still allowing for the presence of permitted 
and semi-forbidden ones \citep{OF06}. For example, for the upper limit the CIII] $\lambda$1909 emission 
line implies that the densities cannot be higher than $\sim$ 10$^{11-13}$ cm$^{-3}$, 
though strong UV lines like Fe II or Al III $\lambda$1860 suggest somewhat higher density, 
at least in the low ionization region \citep{Ba96, La97, Ne11}.

There are not many proposed methods in the literature that
include observations of the BELs to determine the BLR physical
properties \citep{Ma01, La06}. \cite{Ma01} found, using the CLOUDY
photoionization computation, that the ratio of Si III] $\lambda$1892
and C III] $\lambda$1909 UV emission lines is a good density
diagnostic in the density range $9.5 \la \log n_{\rm e} \la 12$. And
since that ratio is correlated with the width of the broad H$\beta$
line, they gave a relationship for the estimates of the electron
density in the BLR using either the ratio of these semi-forbidden
lines or the width of the broad component of the H$\beta$ line
\citep{Ma01}. The same authors extended their analysis and
analyzed the physical conditions using other line ratios 
(e.g. Al III $\lambda$1860/Si III]$\lambda$1892, Si III]$\lambda$1892/C IV$\lambda$1549,
 Mg II$\lambda$2800/Ly$\alpha$, etc.) trying to exploit the CLOUDY simulations
to deduce constraints on ionization parameter, density and column density \citep{Ma10}.
Also, \cite{La06} proposed a method that considers
the electron scattering influence on the line profiles and determines the
physical parameters of the BLR in the case of low luminosity AGN with the 
emission-line profiles having exponential wings (e.g. the case of galaxy NGC 4395).
The method assumes that the exponential wings are produced by the optically
thin, isotropic, thermal electron scattering. In that case, by
fitting the line wings with an electron-scattering model, one can
estimate the electron density and optical depth of the region
\citep{La06}. Both described methods are observationally constrained for
either needing the UV observations or detecting BEL with strong
exponential line wings to extract a direct estimate.

On the other hand, \cite{Po02, Po03, Po06a} suggested that the Boltzmann-plot 
(BP) method (see \S2 for more details), already well known to laboratory diagnostics
of high density plasma \citep{Gr97}, might be exploited to probe the BLR of some AGN 
\citep[see also][]{Il06, LM07, Po08}. By measuring the flux of emission lines 
belonging to a specific transition series, such as the hydrogen Balmer 
line series in the optical domain \citep{Po03}, and using the corresponding atomic parameters, 
the technique provides an estimate of the plasma temperature.
For example, in case of NGC 5548 the BLR temperature was estimated using the BP applied 
to the Balmer lines observed from 1996 to 2004 \citep{Po08}. A high correlation 
between the variation of the optical AGN continuum and the BP temperature
 is found \citep{Po08}. 

The main motivation of this paper is to investigate under which particular
circumstances the Balmer line ratios and BP method may, or may not, be used to explore the physical
conditions of the BLR, given that photoionization is expected to control the plasma physical properties.
To clarify this, we compare the hydrogen Balmer lines (from H$\alpha$ up to 
H$\varepsilon$), that we obtained in three different ways: i) from a grid of 
several numerical models, computed using the CLOUDY spectral synthesis code
\citep{Fe98} and analyzing the properties of the resulting
spectra; ii) by considering the calculated emission line intensities
predicted in the framework of the recombination theory \citep{sh95a}, 
and iii) analyzing a set of observed spectra taken from the Sloan
Digital Sky Survey (SDSS) database \citep{LM07}.

The paper is organized as follows: in \S2 we present the application 
of the Balmer line ratios through the Balmer decrement and BP method, 
in \S3 we describe the grids of models generated with the 
CLOUDY code and report the results of the models, and we describe and analyze the 
theoretical recombination emission line, in \S4 we report on the observed SDSS spectra, 
in \S5 we provide discussion and finally, in \S6 our conclusions are given.

\section{The ratios of the hydrogen Balmer lines - indications of the BLR physical properties}

The hydrogen Balmer lines are the brightest recombination lines in the optical spectra of AGN.
They are usually showing complex line profiles, composed from the narrow and broad component.
Here we discuss only the broad component, coming from the BLR.

\subsection{The Balmer decrement}

The hydrogen Balmer decrement - H$\alpha$/H$\beta$ - ratio can be used to probe
the physics of the line-emitting plasma \citep{KK79, KK81, MBG80, CP81, Do08, Ji12}. 
It is most frequently used to determine the amount of dust extinction
for the low-density gas, such as the narrow line region (NLR) in AGN
\cite{OF06}, where a value of 3.1 for H$\alpha$/H$\beta$ ratio is generally adopted. 
This is slightly higher than the standard case B recombination value as the H$\alpha$ 
emission is enhanced due to collisional effects and harder ionizing continuum 
\cite{GF84, OF06}. 

The densities in the BLR are far higher, as discussed above, so that the
H$\alpha$/H$\beta$ ratio might be affected by
high-density effects or wavelength-dependent
extinction by dust \citep[see e.g.][]{Os84, Go95}.
The observed broad-line H$\alpha$/H$\beta$ ratio of most broad-line AGN 
is usually larger (steeper) than the Case B recombination value  
\citep[see e.g.][]{Os77, Ra85, Do05}, sometimes even as
steep as 10 or higher \citep{Os81, CPW88}. \cite{Do08} obtained the averaged value of 
the broad-line Balmer decrement of 3.06, for a large, homogeneous sample 
of 446 low-redshift (z$\leq$0.35) blue type 1 AGN
with the minimal dust extinction effects. They argue that 
the broad-line H$\alpha$/H$\beta$ ratio can be used as dust extinction 
estimator, even for the BLR, especially for the radio-quiet AGN \citep{Do08}.

On the other hand, in several monitoring campaigns, the Balmer decrement is 
found to vary, usually anti-correlated with the continuum flux,
in a single object \citep{Sh04,Sh10,Po11}, but often exhibiting 
complicated behavior depending on the state of the activity in the AGN. 
In case of NGC 4151, it varied from 2 to 8 during 11 years \citep{Sh08},
usually decreasing when the continuum increases, but staying constant in case
of high continuum flux. A similar effect was detected in the case of 3C 390.3: for the low 
activity period the Balmer decrement is anti-correlating with the continuum,
while in the outbursting phase with higher values of
the continuum flux, it stays almost constant with the value of $\sim$4.5 \citep{Po11}.
Thus the changes in the Balmer decrement might
indicate that sometimes the line production is not fully governed
by the input ionizing continuum coming from the accretion disk,
i.e., the main cause of the Balmer decrement variations is not related to the active nucleus 
and the shock initiated emission is probably dominant \citep{Sh10}.
Or in addition to the physical conditions across the emitting region
or input ionizing flux, the size of the emitting regions
could play an important role \citep{Po11}. 
Concerning the intrinsic host galaxy reddening in
these galaxies, we might expect that during the considered time interval, 
the intrinsic reddening can be neglected, because it should not vary too 
much for such a relatively short period ($\sim$10 years).
Of course, the effects of dust within the BLR might still be present.

As we noted above, the Balmer decrement (only the H$\alpha$/H$\beta$ ratio)
may indicate some physical processes in the BLR, but it is better to take 
into account the ratios of more lines from the Balmer series. One way 
to consider ratios of several lines is to use the Boltzmann-plot. We give a short 
outline of the BP method in the next Section.

\subsection{The Boltzmann-plot}

If a plasma extending over a region of length $\ell$ emits along 
the line of sight, assuming that the temperature and emitter density 
do not vary too much, the flux $F_{ul}$ of a transition from an 
upper level $u$ to the lower $l$ can be calculated as 
\citep{Gr97, K99, Po03, Po06a, Po08}:
$$F_{ul} = \frac{hc}{\lambda} g_{u} A_{ul} \frac{N_0}{Z} \exp (-E_u / kT) \ell, \eqno(1)$$ 
where $\lambda$ is the transition wavelength, $g_u$ is the statistical 
weight of the upper level, $A_{ul}$ is transition probability, $N_0$ is the 
averaged total number density of radiating species which effectively contribute 
to the line flux without being absorbed, $Z$ is the partition function, $E_u$ 
is the energy of the upper level, $T$ is the averaged excitation temperature, 
and $h$, $c$, and $k$ are the Planck constant, the speed of light, and the 
Boltzmann constant, respectively. 
Eq.~(1) additionally assumes that the occupation number of the upper 
level in the transition follows the Saha-Boltzmann distribution, the 
lines are optically thin, and the line series originates in the same emitting 
region \citep[for more detailed derivation and discussion on these 
assumptions see][]{Po03, Po06b, Po08}. 

Introducing a ``normalized'' line-flux $F_{\rm n}$, with respect to the 
atomic constants which characterize the transition:
$$F_{\rm n}={F_{ul}\cdot \lambda\over{g_u A_{ul}}}, \eqno(2)$$
we see from Eq.~(1) that it is possible to write:
$$\log_{10} (F_{\rm n}) = B - A E_u, \eqno(3)$$
where, considering lines that belong to a specific transition series, 
$B$ and $A$ represent the Boltzmann-plot parameters, with $B$ being a characteristic 
constant of the transition series and $A = \log_{10}(e)/kT_{\rm exc} \approx 5040 / T_{\rm exc}$ 
is called the temperature parameter. 

Therefore, taking into account a specific line series, such as the Balmer lines, 
if the populations of the upper energy states ($n \geq 3$)\footnote{Since 
the emission deexcitation goes as $u\to l$ it is not
necessary that level $l$ has the Saha-Boltzmann distribution.} follow a 
Saha-Boltzmann distribution, Eq.~(3) provides an 
estimate of the average excitation temperature in the 
region where the lines are formed. In the case of a low density environment, 
where the narrow lines come from, the above assumptions do not apply 
\cite[see][]{Po03}, but for higher densities in the BLR plasma, 
the approximation might be reasonable. The Balmer line ratios are less 
sensitive to temperature changes at higher temperatures \citep{Gr97}. 
For further analysis we adopt the maximal value
of the BP temperature to be 30,000 K, as the BLR temperature should be 
lower than that \citep{OF06}.

\begin{figure}
\centering
\includegraphics[width=8.0cm]{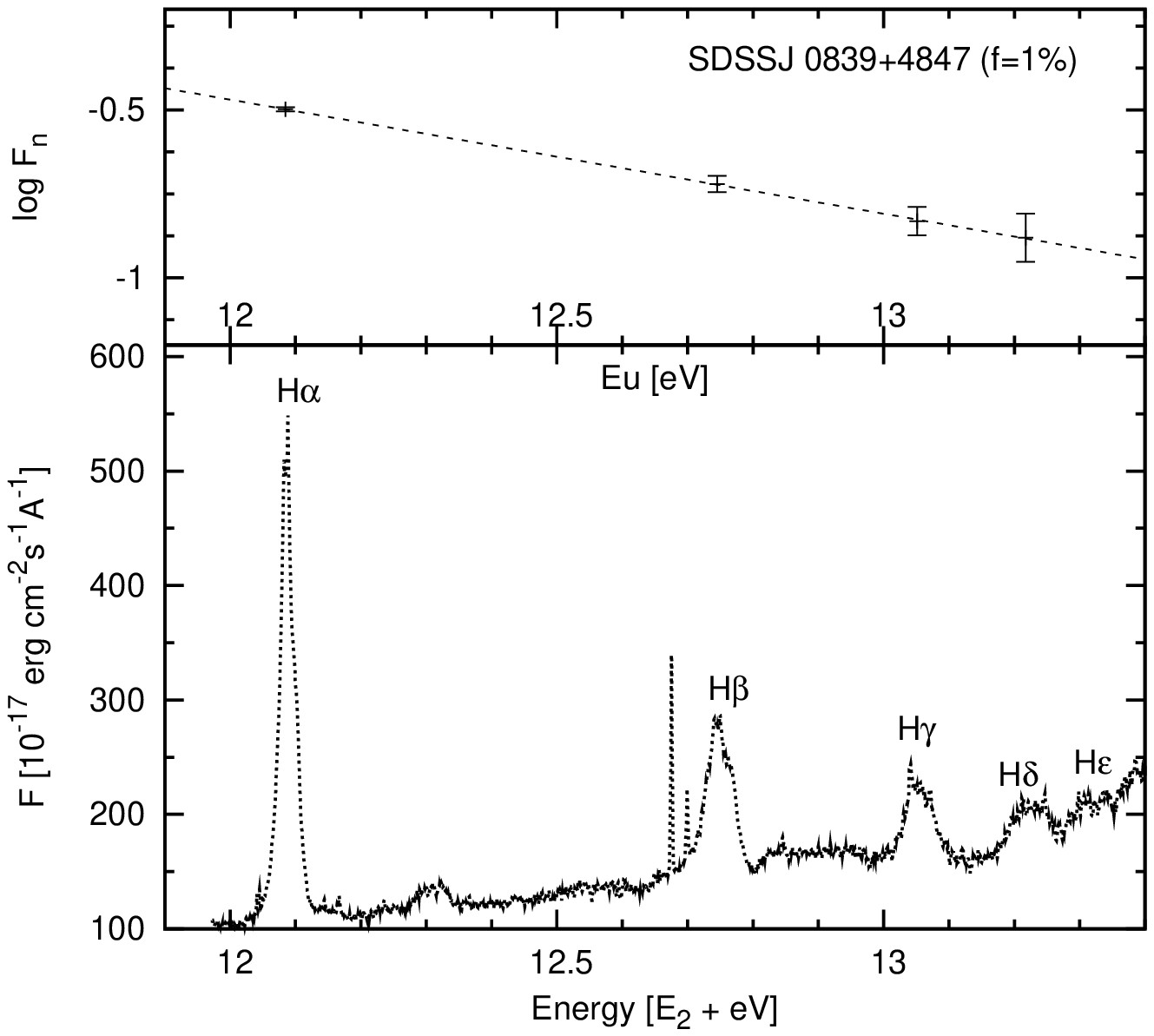}
\includegraphics[width=8.0cm]{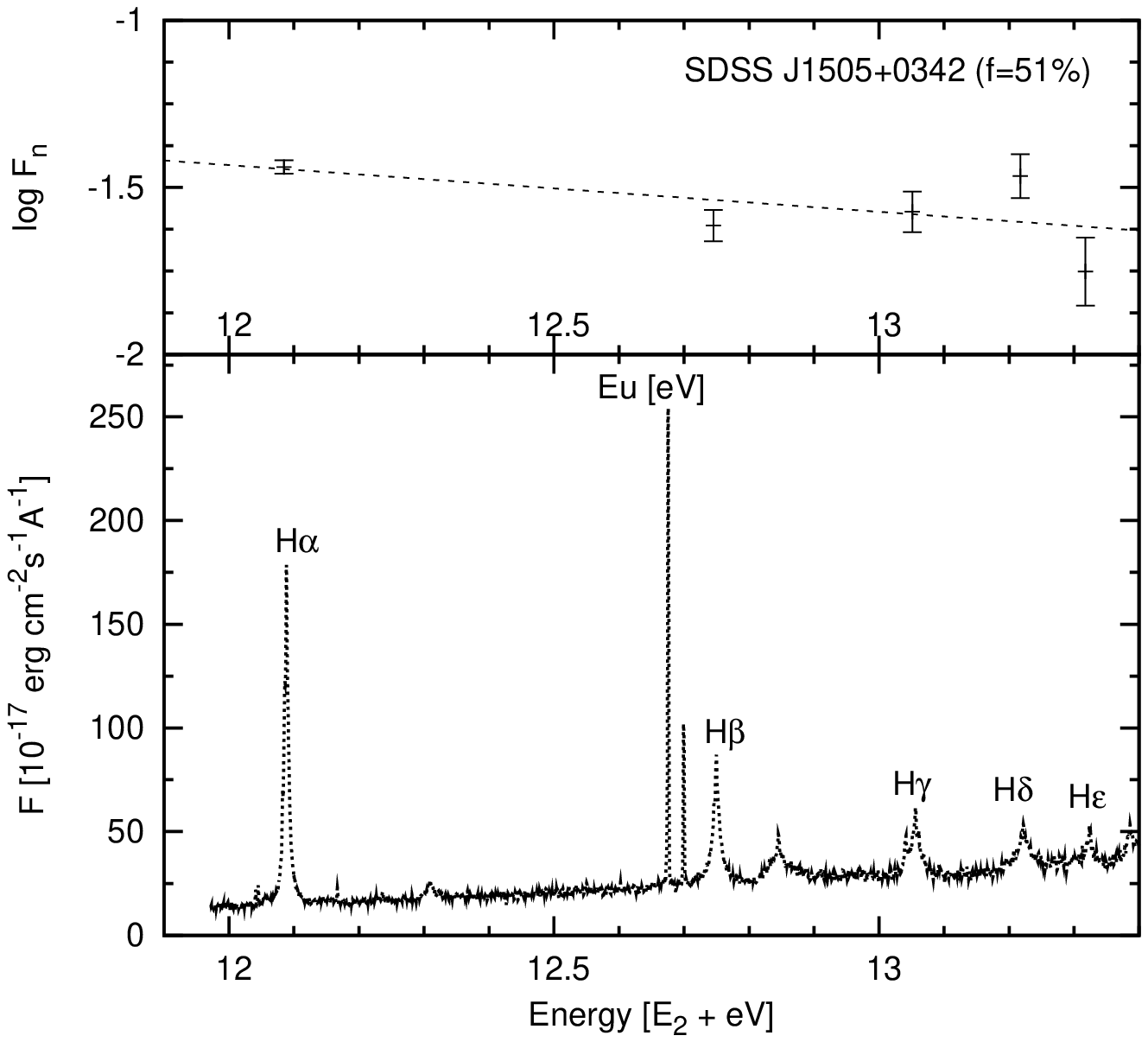}
 \caption{Examples of AGN where the BP method works (upper panel)
and where it does not give valid results (bottom panel) applied on 
the data taken from \citet{LM07}. Each panel contains the observed spectrum and the
BP applied on the line fluxes of the Balmer series (only the broad component
of the Balmer lines was used). For the lower plots, which give the observed spectra, 
the observed wavelength scale (x-axis) is converted into energy units (eV), 
according to the energy above the ground of the upper transition levels, to match the BP 
in the upper plots.}
\label{fig1}
\end{figure}

The Boltzmann-plot should not be confused with the Balmer decrement,
as it considers line-ratio normalized using the atomic
parameters of the specific line transitions. However, all 
discussion presented for the Balmer decrement in the previous section, 
applies for the BP since the line ratios could be highly influenced 
by the dust effects. On the other hand, the BP method has the clear advantage to require only the measured 
Balmer line fluxes to estimate the excitation temperature. But one has to consider some 
possible drawbacks, concerning the use of emission lines to infer the BLR 
physical properties, which in general appear in all methods that use BEL parameters
for plasma diagnostics. Since the BEL profiles are complex, the estimate of 
their fluxes should deal with the possibility of a multiple-component structure 
in the BLR, as stated in previous section. Furthermore, there are some indications suggesting that the 
Balmer lines do not necessarily have to arise strictly from the same region, 
as it is pointed out by some differences in the profiles of H$\alpha$ and H$\beta$ 
\citep[see e.g][]{Sh08, Po11}. Finally, different mechanisms may contribute to 
the formation of the emission lines and, though photoionization is the main heating 
source, other processes might be relevant. The combination of these effects implies 
that the actual properties of the BLR may or may not fit in the fundamental assumptions 
of the BP method, as it is illustrated in the examples plotted in Fig.~\ref{fig1}.

\begin{figure}[]
\centering
\includegraphics[width=4.2cm,angle=-90]{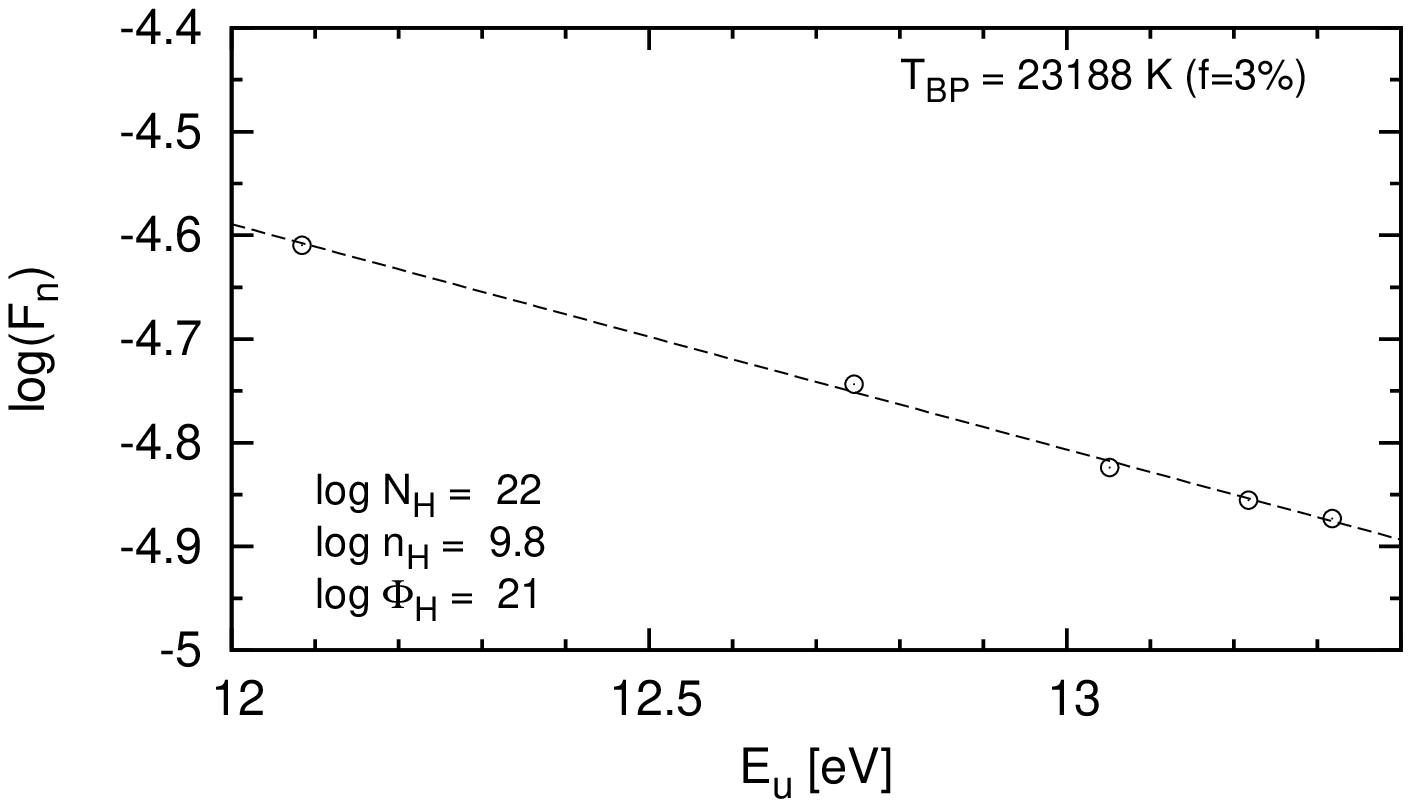}
\includegraphics[width=4.2cm,angle=-90]{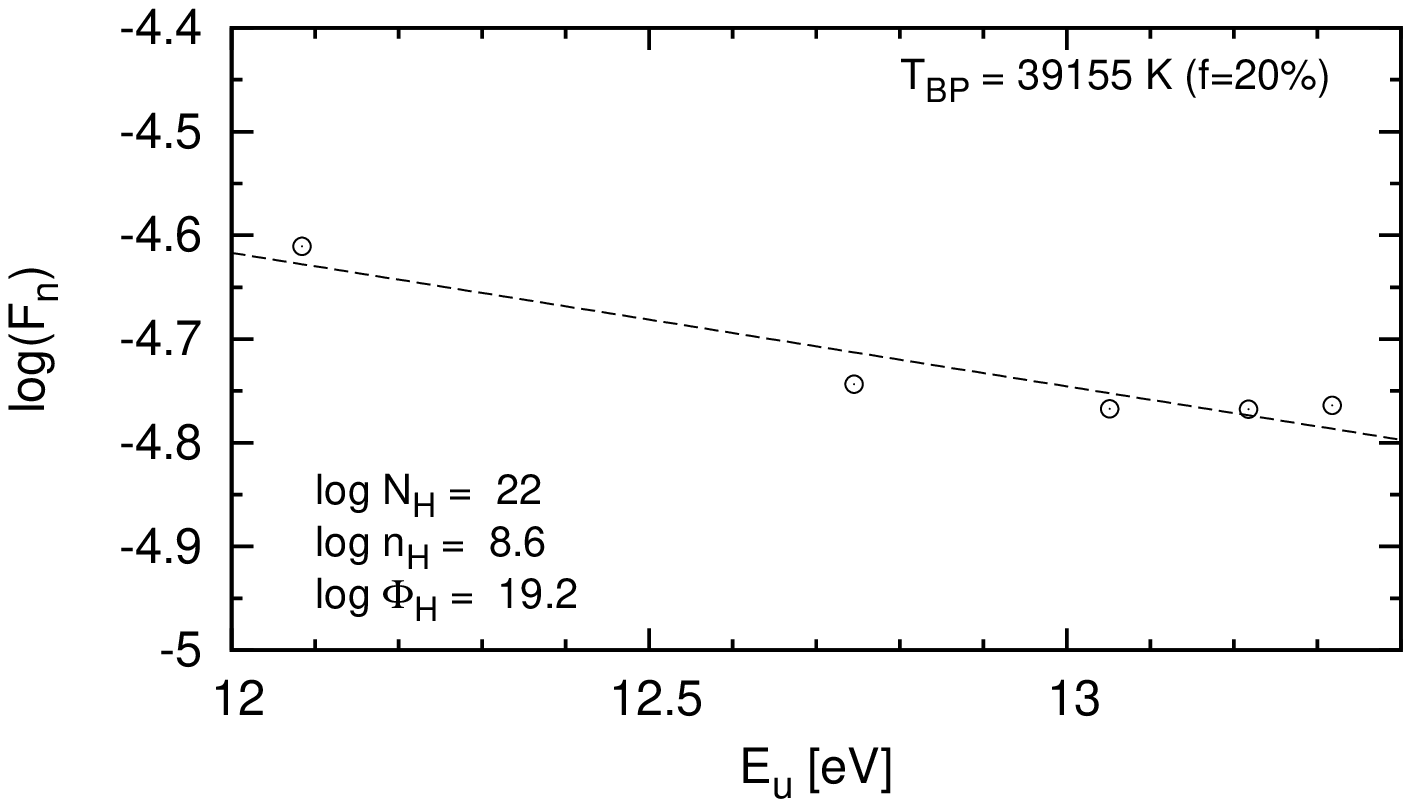}
\includegraphics[width=4.2cm,angle=-90]{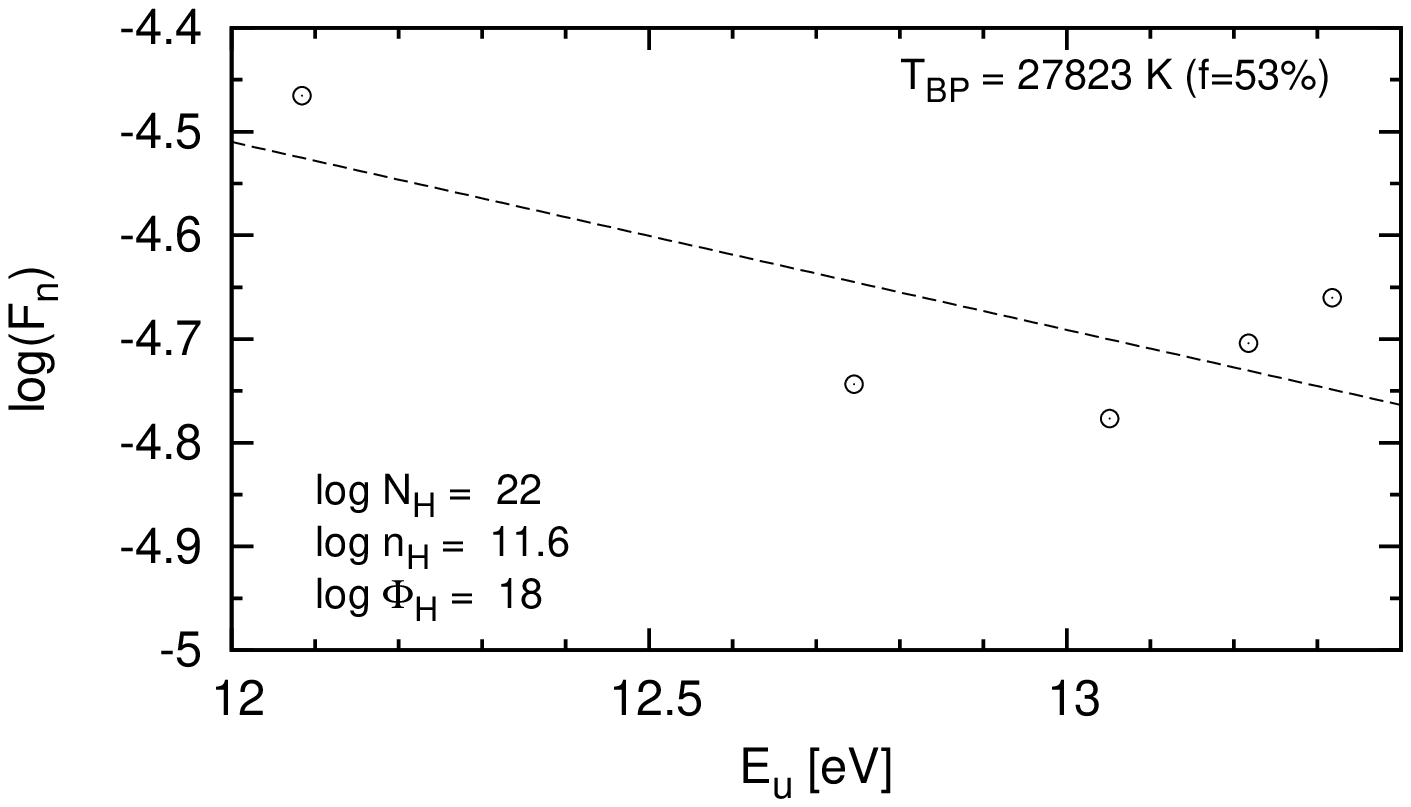}
\caption{Different examples of the BP applied to the Balmer lines
obtained with the CLOUDY models. The intensities $F_{\rm n}$
are calculated using the Balmer lines normalized to the H$\beta$
flux. In the upper right corner, the BP temperature $T_{\rm
BP}$ and the error of the BP fit $f$ (in percentage) are given. 
In the bottom left corner the column density $N_{\rm H}$, 
the hydrogen density $n_{\rm H}$ and 
the input ionizing flux $\Phi_{\rm H}$ are given.} \label{fig2}
\end{figure}

\section{Simulated and theoretical Balmer line ratios}

\subsection{Simulation of photoionized region and modeled Balmer lines}

Numerical simulations make it possible to understand complex physical 
environments starting from fundamental principles of physics. This is 
particularly applicable in the case of the BLR where due to different 
physical conditions in the plasma (especially high density), many
physical processes should be taken into account. To model such a complex 
physical scenario, our choice was to use CLOUDY\footnote{All details related 
to this code can be found at: {\it http://www.nublado.org/}} \citep{Fe98}. 
Collisional effects, including excitation and de-excitation, continuum 
fluorescence, recombination, etc., are all accounted for in this code \citep{Fe98,Fe06}.

\begin{figure*}[]
\centering
  \includegraphics[width=6cm]{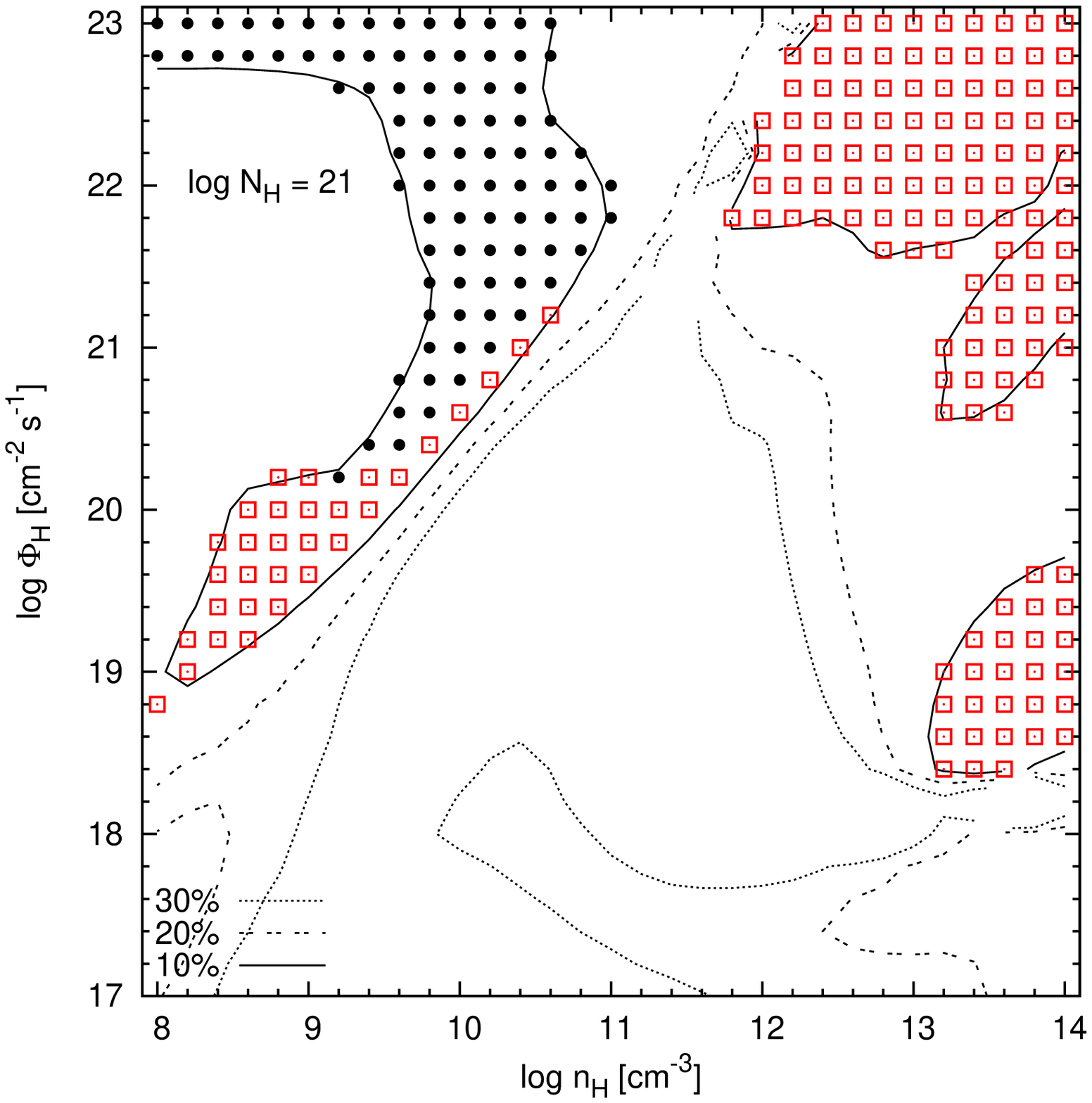}
  \includegraphics[width=6cm]{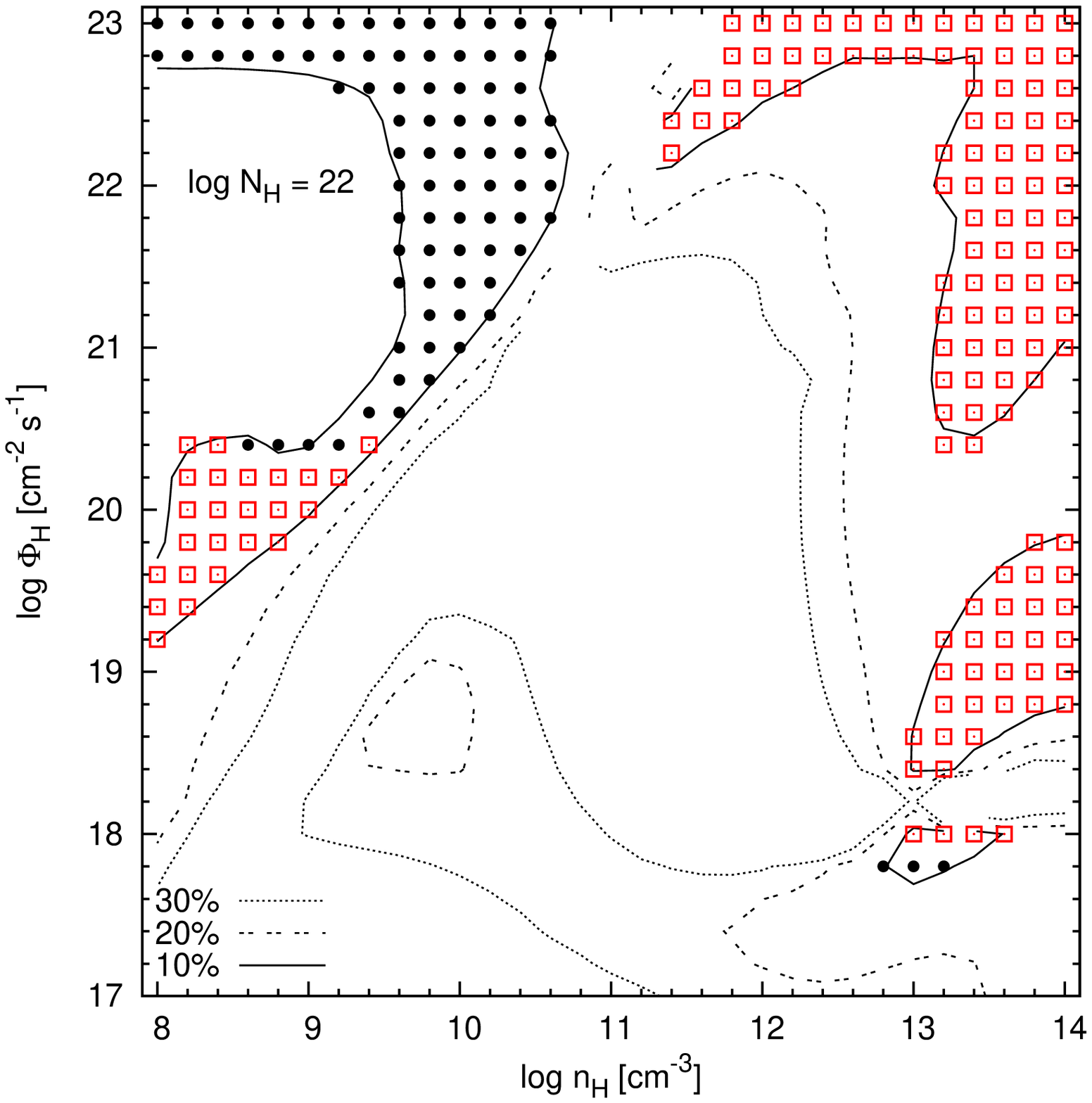}
  \includegraphics[width=6cm]{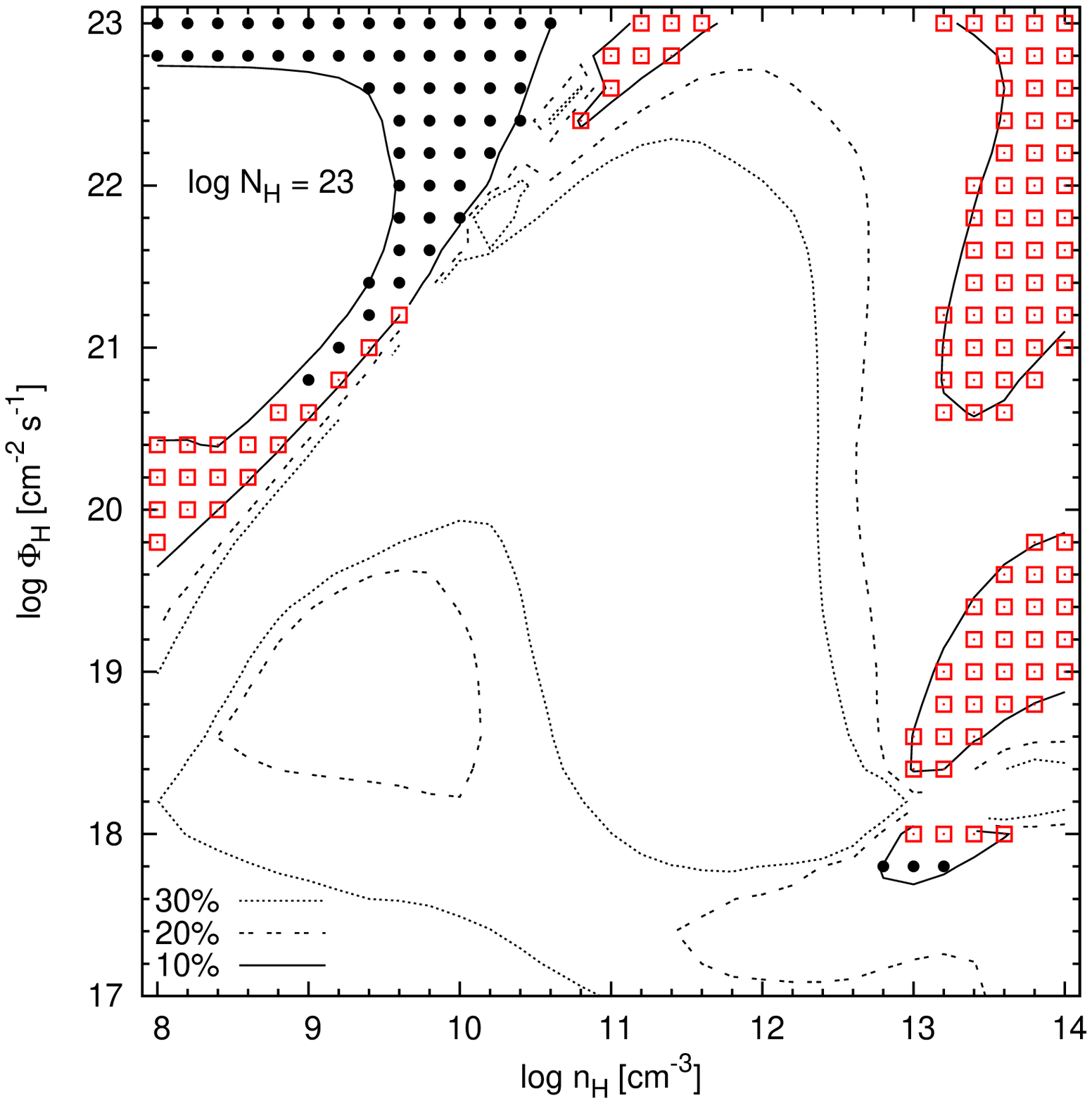}
  \caption{Contour plots for the error of the BP fit $f$ in the ionizing 
flux and hydrogen density plane for different column densities $N_{\rm H} \in [10^{21} - 10^{23}] 
\ {\rm cm^{-2}}$. The curves show the regions where $f$ is smaller than 10\%, 20\% and 30\%, while 
open squares represent the simulations for which $f\leq10\%$, and full circles are the simulations
for which $f\leq10\%$ and $T_{\rm BP}<30,000$ K.}\label{fig3}
\end{figure*}

In order to investigate the Balmer lines, we generate grids of 
photoionization models of the BLR using the version C08.00 of CLOUDY
\citep[][]{Fe98, Fe06}. This version is roughly a factor of 2 
slower than C07.02, due to a major expansion in the physics of the H-like 
iso-sequence. The l-levels of the H-like sequence are now fully resolved,
making these atoms formally correct, thus the predicted H~I emission line 
spectrum is changed with respect to the C07.02. Input parameters for the 
simulations are chosen to match the standard conditions in the BLR 
\citep{Fe06, KG00, KG04}. Assuming to have solar chemical abundances 
and constant hydrogen density, and using the 
code's AGN template for the incident continuum shape (a continuum similar to 
the typical radio quiet active galaxies), we compute an emission line 
spectrum for the coordinate pair of hydrogen gas density $n_{\rm H} [{\rm cm}^{-3}]$ 
and hydrogen-ionizing photon flux $\Phi_{\rm H}[{\rm cm^{-2}s^{-1}}]$.

The grid dimensions span 6 orders of magnitude in each direction and, 
with an origin in $\log n_{\rm H} = 8$, $\log \Phi_{\rm H} = 17$ and a step of 0.2 dex 
increments, produced a total of 961 simulations per grid. The column density $N_{\rm H}[{\rm cm}^{-2}]$ is 
kept constant in producing the grid of simulations. Many authors claim that the most 
probable value of the column density in the BLR is $N_{\rm H} = 10^{23}{\rm cm}^{-2}$ 
\citep{Du98, KG00, KG04}. Since the column density defines the optical thickness of 
the region, here we produce 3 grids of models changing the column density 
between $N_{\rm H} = [10^{21}, 10^{22}, 10^{23}]{\rm cm}^{-2}$. 
Therefore, the total number of simulations is 2883.

The output of the code includes all lines\footnote{By default, the code gives line 
fluxes normalized to the H$\beta$ flux. Since it has no influence in the BP analysis, 
we have used the normalized values.} formed in the emitting region, with a 
given chemical composition, hydrogen density, and hydrogen-ionizing photon 
flux. We consider in our analysis the strongest hydrogen Balmer lines 
(H$\alpha$ to H$\varepsilon$).

\subsubsection{Modeled line ratios}

For every grid of CLOUDY models, we analyze the emission line fluxes 
by applying the BP method to the Balmer lines. In particular, using Eq. (3), 
we estimate the parameter $A$, from which we then calculate the BP temperature 
(hereinafter denoted as $T_{\rm BP}$) of the emitting region and the error associated 
to the best fit in the BP, which we denote as $f$.

A few examples of the BP applied on the simulated Balmer lines for a column density 
of $N_{\rm H} = 10^{22}{\rm cm}^{-2}$ are presented in Fig.~\ref{fig2}. In most of the cases 
a satisfactory fit of the Eq. (3) is not obtained and $f$ has pretty large values. 
This is more noticeable in the plots of $f$ in the hydrogen density and ionizing flux 
plane for all 3 grids of models, illustrated in Fig.~\ref{fig3}. 
It can be seen in Fig.~\ref{fig3} that the input parameters $\Phi_{\rm H}$, $n_{\rm H}$ 
for which $f$ was smaller than 10\% are well constrained (open squares in Fig.~\ref{fig3}), 
and in a similar range for different column densities. 
The realistic models for which $T_{\rm BP}<30,000$ K (denoted with full circles 
in Fig.~\ref{fig3}) have a very high ionization parameter (low density and high input
ionization flux), except for a couple of cases with high densities
appearing at higher column densities.

\subsection{Theoretical line ratios}

We consider also theoretical recombination Balmer lines calculated 
by \cite{hs87, hs92, sh88, sh95a}. We want to analyze the hydrogen 
Balmer lines obtained only by using the recombination theory, and compare 
them to the line ratios obtained by CLOUDY code, which, on the other hand,
treats simultaneously millions of lines and physical processes, so that the  
behavior of the hydrogen lines might be hidden. 

\cite{sh95a} calculated line emissivities, together 
with other line transition parameters for hydrogenic atoms, assuming both Case A
and Case B recombination, and including full collisional effects for a considerably
large range of temperature, density, and principal quantum numbers. Their
calculated data are given in tables available in an online 
catalogue\footnote{http://vizier.cfa.harvard.edu/viz-bin/VizieR?-source=VI/64}
\cite[see][for details]{sh95a, sh95b}

Here we use their theoretical hydrogen Balmer lines (H$\alpha$ to H$\varepsilon$),
calculated for the Case B recombination with the density $\log n_{\rm e}$ in the 
range (8.2--14, 0.2 dex increment), and the electron temperature lying within the 
range 1000 to 30,000 K (step 1,000 K). Online data provide line emissivities within 
this range, but only for a small number of values of density and temperature  \citep{sh95a},
for others we interpolated line emissivities using the interpolation
code provided by \cite{sh95a}. For the BP method we use the ratios of Balmer lines,
which we could directly derive from line emissivities.

\begin{figure}
\centering
  \includegraphics[width=8.5cm]{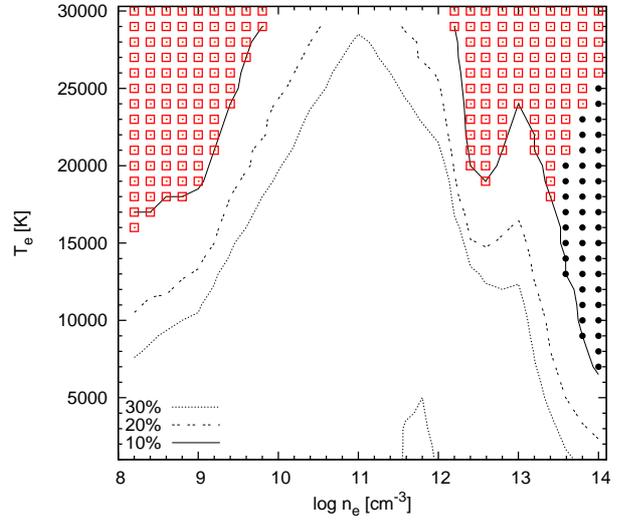}
  \caption{Contour plot for the error of the BP fit $f$ in the electron
temperature vs. electron density plane for theoretical Balmer line ratios. 
The curves show the regions where $f$ is smaller than 10\%, 20\% and 30\%,
while points have the same meaning as in Fig. \ref{fig3}.}\label{sh}
\end{figure}

\begin{figure}
\centering
  \includegraphics[width=8.7cm]{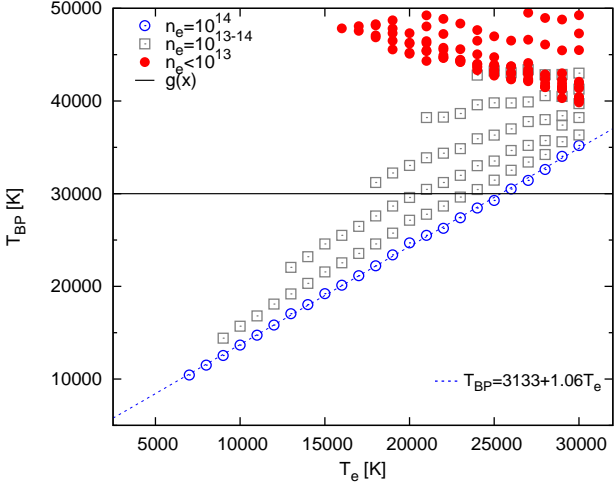}
  \caption{BP temperature plotted against the electron temperature for the case
when $f\leq10\%$. Open circles are for the electron density of $10^{14} \rm cm^{-3}$,
open squares for $10^{13}\leq n_{\rm e} < 10^{14} \rm cm^{-3}$ , and full 
circles for $n_{\rm e} <10^{13} \rm cm^{-3}$. The linear best-fitting
of the cases with $n_{\rm e}=10^{14} \rm cm^{-3}$ is given with the dashed line. 
The solid line marks the $T_{\rm BP}=30,000$ K. }\label{tt}
\end{figure}

\begin{figure}
\centering
  \includegraphics[width=8.1cm]{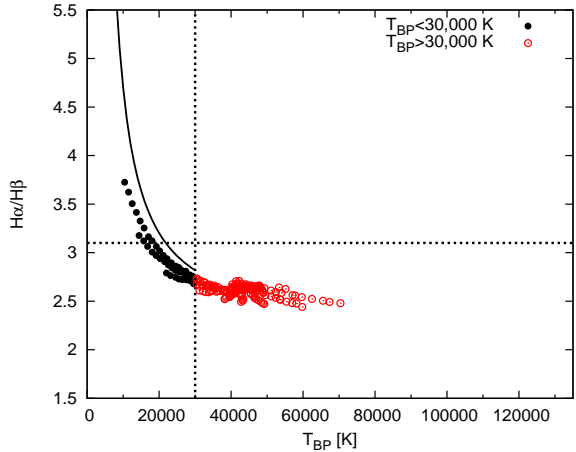}
  \caption{H$\alpha$/H$\beta$ ratio plotted against the BP temperature for 
 $f\leq10\%$, where full circles denotes $T_{\rm BP}<30,000$ K and
open circles $T_{\rm BP}>30,000$ K, and are separated with the vertical dashed line.
The horizontal dashed line  represents ${\rm H}\alpha/{\rm H}\beta=3.1$. 
The solid line gives the line ratio calculated using Eq.(1).}\label{HabT}
\end{figure}

We plot the error of the BP fit $f$ for theoretical Balmer line ratios 
in the electron temperature and electron density plane (Fig.~\ref{sh}). 
Here we do not have the input ionizing flux which is used as an input parameter 
for modeled line ratios. But we can say that basically the electron temperature  
gives some indications about the input ionizing flux. 
For the theoretical line ratios, we obtain that the $f\leq10\%$ 
area is also covering a constrained area of parameter space (open squares in Fig.~\ref{sh}), 
which is resembling the one obtained from the simulated line ratios.
Here the cases with $T_{\rm BP}<30,000$ K (denoted with full circles 
in Fig.~\ref{sh}) appear for high densities $\sim 10^{13-14} \rm cm^{-3}$.

The BP temperature as a function of the electron temperature
is plotted in Fig.~\ref{tt}, for cases when the BP fit error is less than 10\%. 
Clearly there is a relation between these two temperatures, especially for the
electron density of $10^{14} \rm cm^{-3}$ (open circles in Fig.~\ref{tt}).
The BP temperature, that represents the excitation temperature is higher than 
the electron temperature. The linear best-fitting gives the following relation
$T_{\rm BP}= 3133 + 1.06 \times T_{\rm e}$. 

The Balmer decrement when $f\leq10\%$ follows the behavior of the Boltzmann-plot 
(solid line in Fig. \ref{HabT}).
For these cases in Fig. \ref{HabT} the H$\alpha$/H$\beta$ ratio is shown against 
the BP temperature. The solid line represents the Balmer decrement calculated 
for different temperatures using Eq.(1), and it is shown only for temperatures 
smaller than 30,000 K. The averaged H$\alpha$/H$\beta$ is 2.96 for cases 
when  $T_{\rm BP}<30,000$ K (full circles in Fig. \ref{HabT}).

\begin{figure}
\centering
  \includegraphics[width=8.5cm]{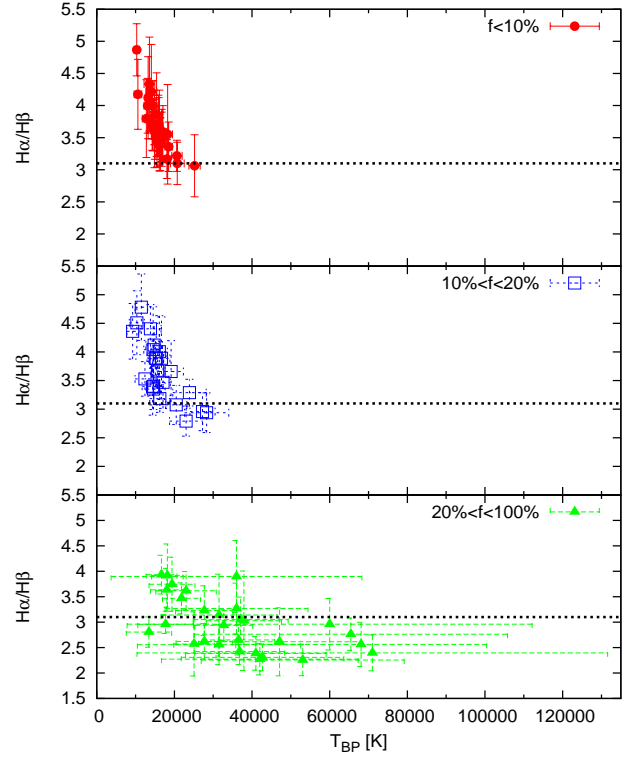}
  \caption{H$\alpha$/H$\beta$ ratio vs. the BP
temperature for the sample of SDSS galaxies for
objects that have the BP fit error $f\leq10\%$ (upper panel, full circles), 
objects with $10\%<f\leq20\%$ (middle panel, open squares), 
and with $20\%<f\leq100\%$ (bottom panel, full triangles) are given. 
The dashed line represents ${\rm H}\alpha/{\rm H}\beta=3.1$. }\label{fig9}
\end{figure}

\section{Observed line ratios}

We explore the usage of the BP method on the observed data. For that we consider 
measurements performed on a sample of 90 broad-line-emitting AGN, taken from 
the SDSS spectral database. For this sample the Balmer line parameters were 
accurately estimated considering the complexity of BEL \citep[see][for details]{LM07}.
We repeat  here that the line fluxes are corrected for the Galactic extinction, as well
as for the host galaxy contribution, which is to some extent correcting for 
the internal dust extinction, but not for the dust effects within the BLR. 
The BP method, applied to the broad component of Balmer lines of this sample 
(Fig.~\ref{fig1}), produces reasonably good fit uncertainties, 
with $f\leq10\%$, in approximately 35\%\ of cases. The BP temperatures are in the 
range $T_{\rm BP}=10,300-25,200$~K.
We used these cases for which $f\leq10\%$ to calculate $T_{\rm e}$ assuming that the
electron density is $n_{\rm e}=10^{14} \rm cm^{-3}$ and using the linear relation from Fig. \ref{tt}. 
We obtain the electron temperature in the range $T_{\rm e}=7,350-22,250$~K. The
obtained values for both temperatures are in a good agreement with the values 
considered to be typical for the BLR plasma.

We plot the H$\alpha$/H$\beta$ ratio against the BP temperature for this sample of 
90 SDSS galaxies (Fig. \ref{fig9}) denoting differently objects with different $f$. 
Objects for which the BP method is working (full circles, upper panel Fig. \ref{fig9}) 
tend to have higher H$\alpha$/H$\beta$ ratio (averaged value is 3.68), in contrast to objects with high fitting 
error $f$ (full triangles, bottom panel \ref{fig9}) that mostly show low Balmer decrements,
below 3.1 (dashed line, Fig. \ref{fig9}).

\begin{figure*}
\centering
  \includegraphics[width=6cm]{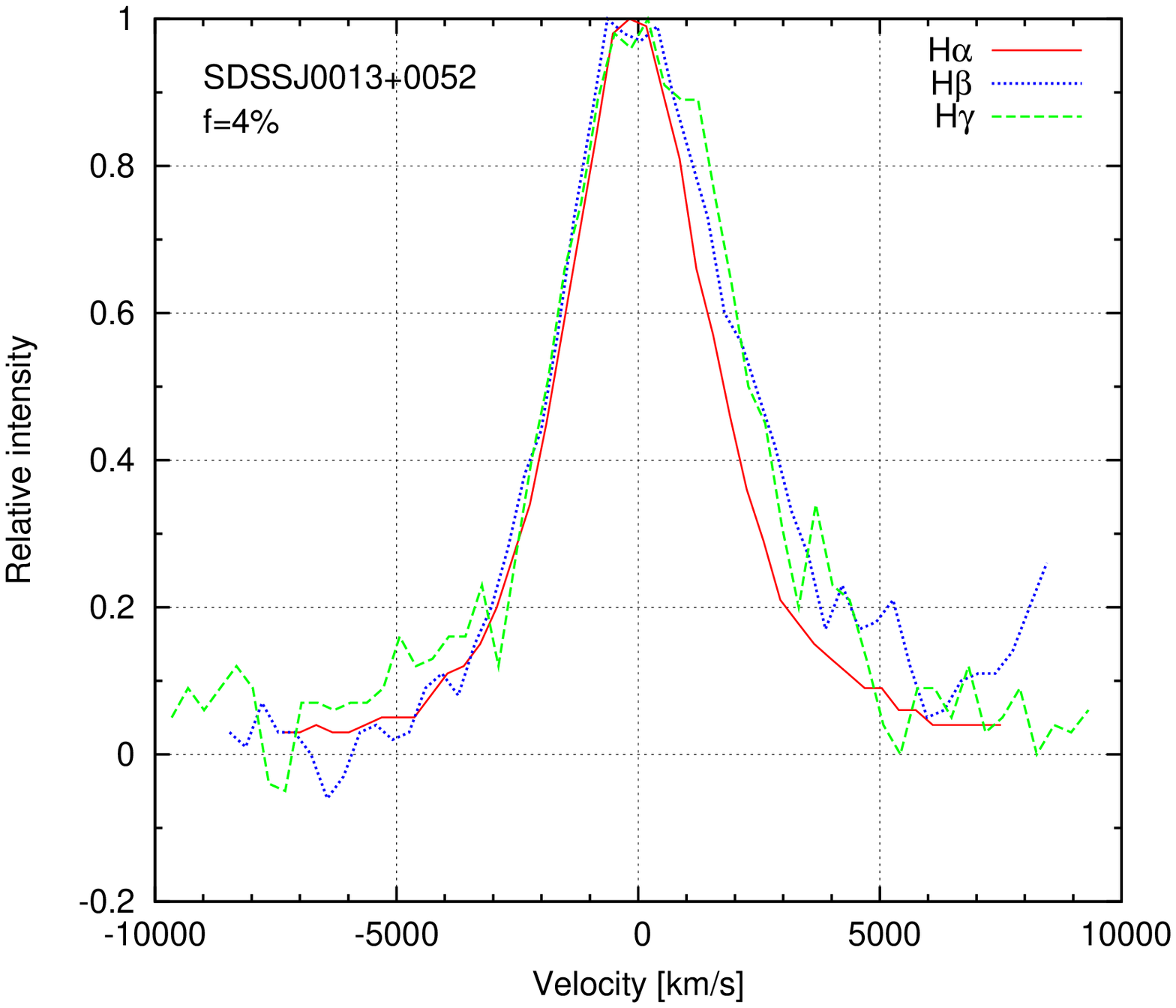}
  \includegraphics[width=6cm]{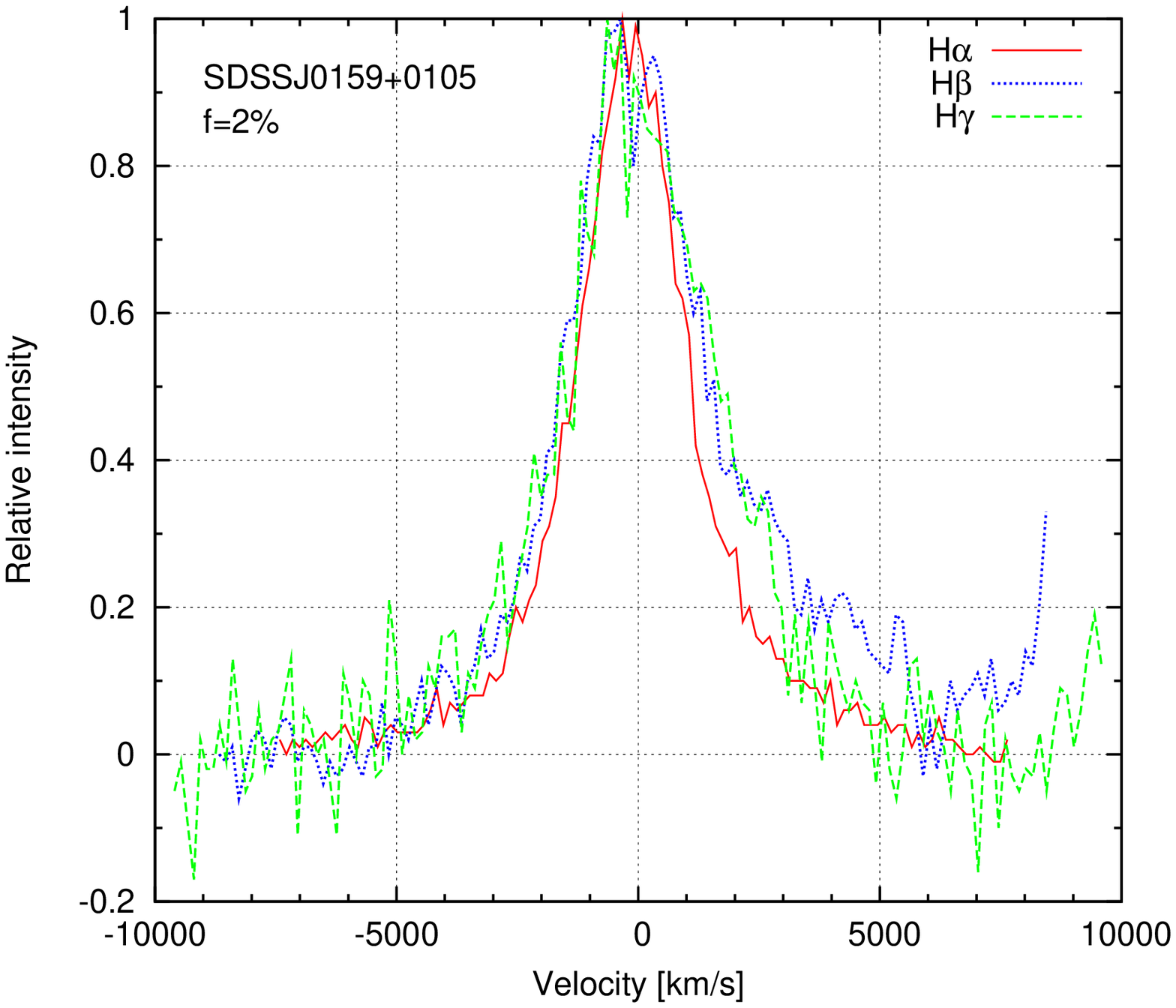}
  \includegraphics[width=6cm]{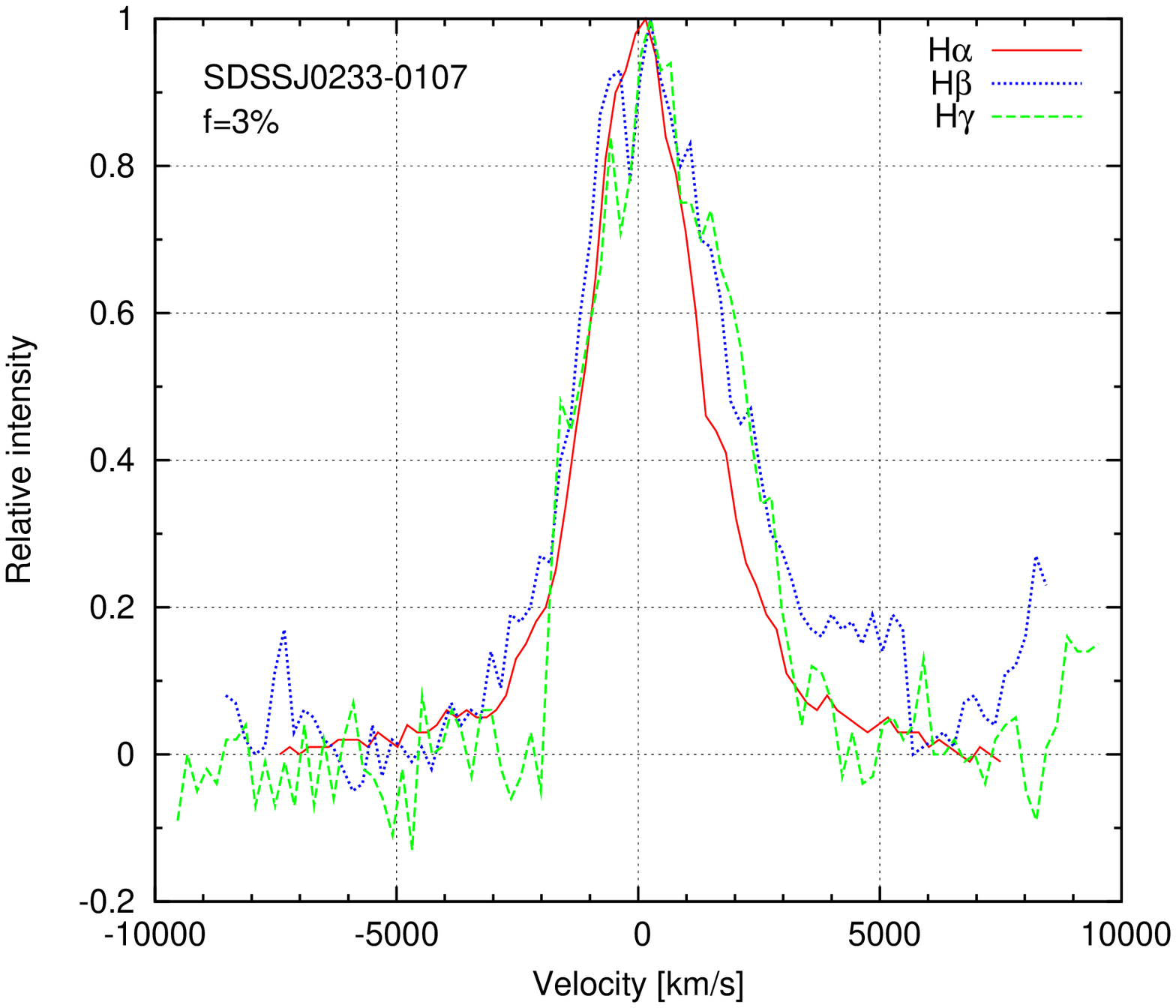}
  \includegraphics[width=6cm]{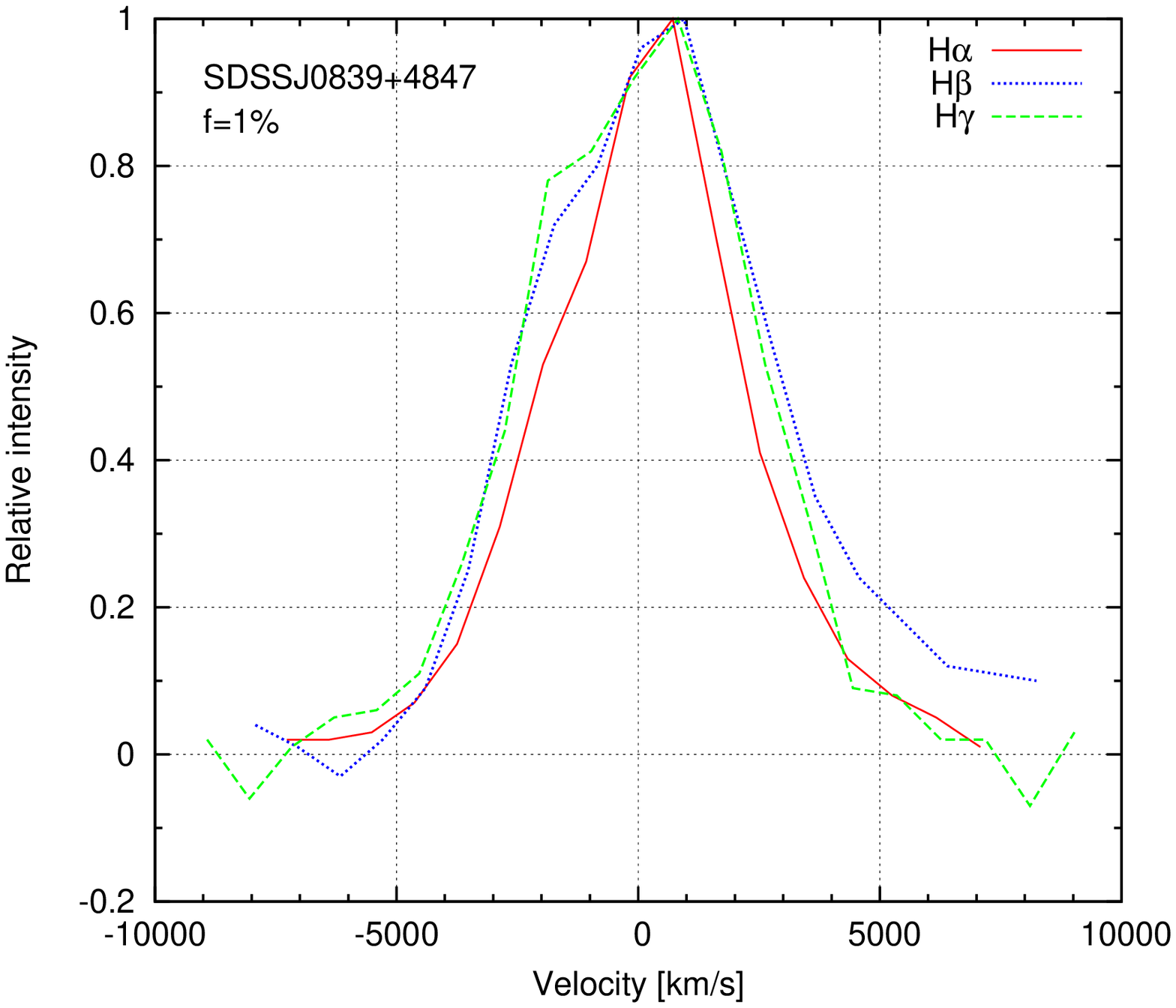}
  \includegraphics[width=6cm]{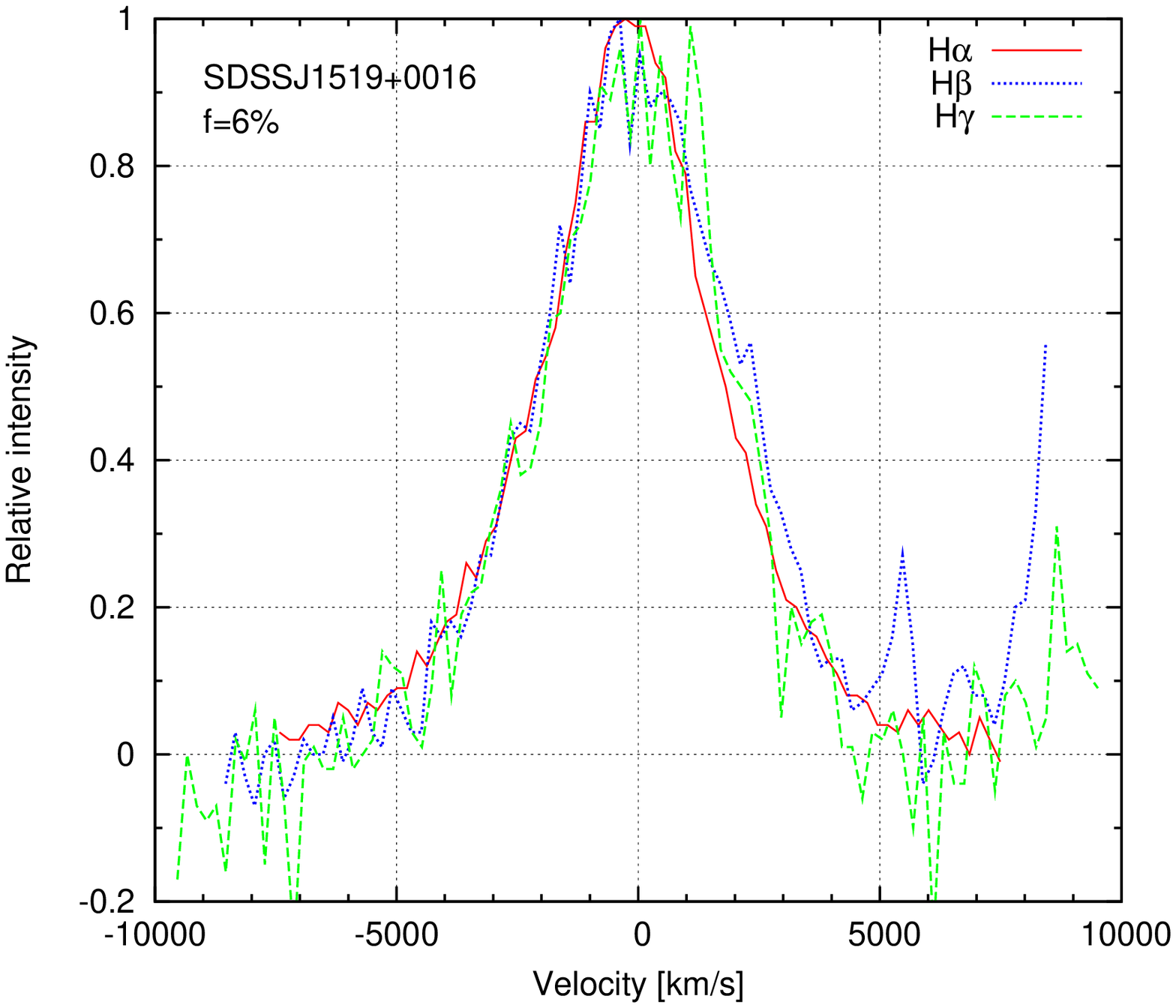}
  \includegraphics[width=6cm]{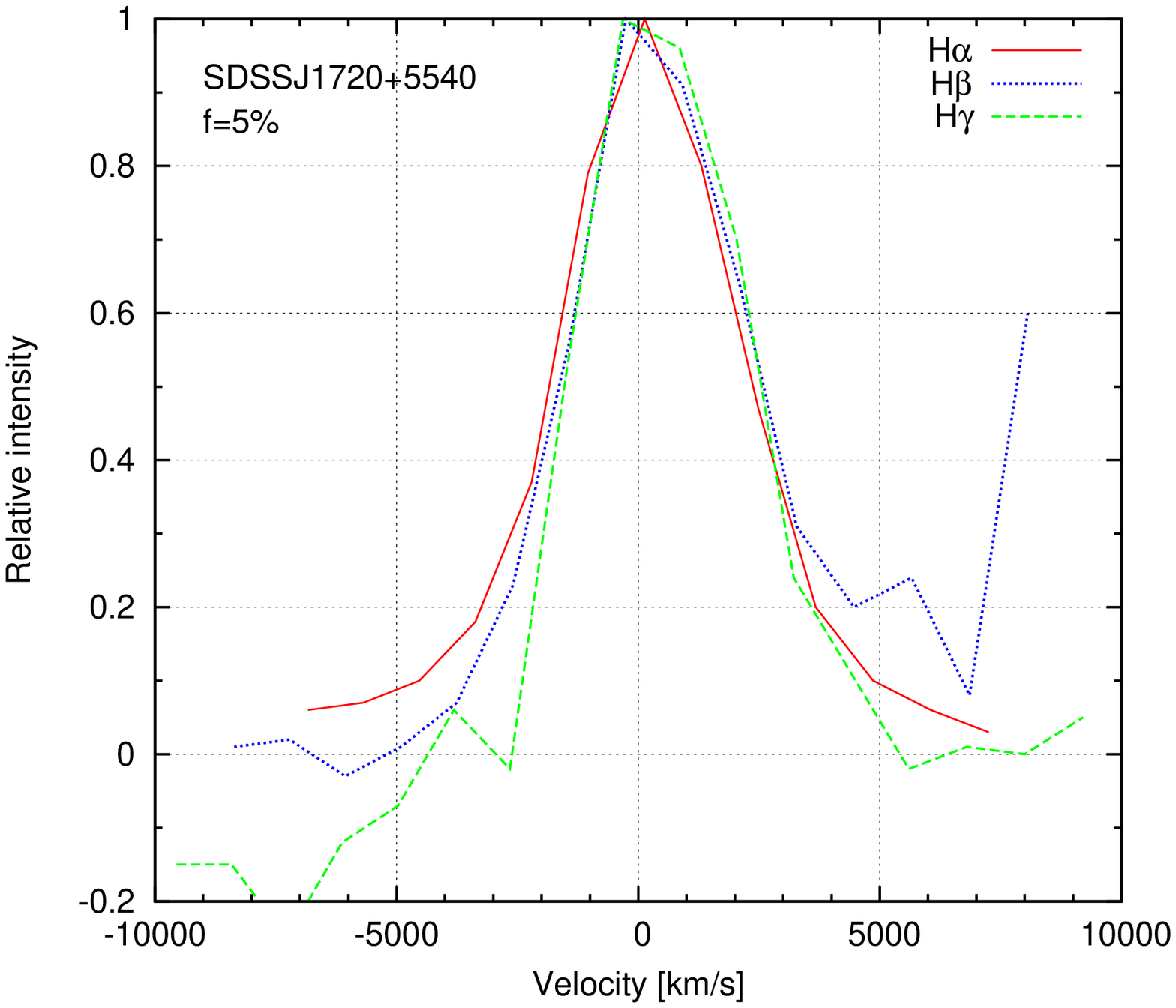}
  \caption{Some examples of the line-profiles comparison for the 
SDSS galaxies for which the BP fit error is less than $10\%$.
The H$\alpha$ (solid line), H$\beta$ (dotted line), and H$\gamma$ 
(dashed line) broad components are shown. The name of the SDSS galaxy and the 
BP error $f$ are given. Lines are smoothed according
to the spectral S/N ratio, normalized to
unity, and converted to the velocity scale.}\label{sm}
\end{figure*}

One indication of the Balmer lines optical depths can
be seen if we compare their normalized-line profiles.
 Some examples of the line-profiles comparison from the SDSS sample 
of galaxies for which $f\leq10\%$ is plotted in 
Fig. \ref{sm}. The H$\alpha$, H$\beta$ and H$\gamma$ 
line profiles (only the broad component) are compared after being smoothed according
to their spectral S/N ratio and normalized to unity.
From the sample of 90 objects, objects with small BP error $f$ 
contain H$\alpha$, H$\beta$ and H$\gamma$ lines which are 
showing similar line profiles (Fig. \ref{sm}), 
indicating that these lines  may be optically thin. One should have in mind 
the relatively low resolution of the SDSS, which is especially seen in the
case of lines with low FWHM, whose line profiles are significantly
affected by this.

\section{Discussion}

According to the simulation results there is a constrained parameter space, 
at different column densities $N_{\rm H} = 10^{21} - 10^{23} \ {\rm cm^{-2}}$, where 
the BP produces reasonable results with small fit uncertainties ($f\leq10\%$). 
Here the cases with $T_{\rm BP}<30,000$ K (denoted with full circles, Fig. \ref{fig3})
that represent less than 9\% of total number of simulations, are mostly falling within an 
area with higher ionizing fluxes and lower hydrogen densities, indicating high
ionization parameters. These situations are extreme ones, thus the most favorable
cases for the BP should happen for higher density, as it is seen in the case
of higher column densities (Fig.~\ref{fig3}). 

A similar trend is seen in case of the theoretical line profiles (Fig.~\ref{sh}). 
Here the parameter range where the BP method is working is constrained and 
higher densities are preferred $\sim10^{14} \rm cm^{-3}$ (Fig.~\ref{tt}). 
In these situations, even if photoionization still acts as the main 
heating source for the BLR, a contribution of the collisional processes 
 may become relevant (especially for such high densities) and significantly 
affect the distribution of the excited energy configuration of the ions. 
The BP method reveals the occurrence of such 
circumstances and it provides an estimate of the excitation temperature,
which basically describes the Boltzmann equation of the
hydrogen energy levels.

In the sample of 90 broad-line-emitting sources from SDSS, there are 
more than 30\% cases for which the BP can be applied and it gives 
reasonably good fit uncertainties ($f\leq10\%$), we denote 
these object as the ``BP objects''. From the observations we see that 
the BP objects have higher Balmer decrement (Fig. \ref{fig9}),
with the average value of H$\alpha$/H$\beta$=3.68. This value is higher than the 
averaged value of 3.06 of the broad-line Balmer decrement of \cite{Do08}
obtained for blue type 1 AGN. Our measured Balmer decrement can indicate two situations: 
(i) either there is higher contribution of collisional processes, i.e. we have high 
density plasma in the BLR (that is also supported by the recombination theory and 
modeled line ratios) which has been suggested in some earlier works  
\citep{vG87, ST93, Br94},
(ii) or that such high H$\alpha$/H$\beta$ ratio is caused by 
the intrinsic dust extinction effects (see discussion in \S 2.1). On the other hand, the obtained
Balmer decrements and the temperature parameter $A$ (see Eq.1) of the SDSS  objects 
that follow the BP are not preferring high continuum luminosities 
\citep[see Fig. 11 in][]{LM07}. If we put this in the context of the modeling and theory
findings, we  can say that these objects then tend to have higher density in the BLR.

We further discuss the possible mechanism of line production in such BLR plasma,
especially the requirements to have optically thin Balmer lines. 
We start from the basic assumption that we have higher opacities in the 
Lyman line series while the Balmer line series is optically thin. In such a case, the 
Ly$\alpha$ line is scattered, while the Ly$\beta$ is almost completely suppressed. 
Therefore, we constantly have atoms excited to the energy level n=2. Since the energy 
gap between the n=2 and n=3 level is 1.9 eV, corresponding to an electron 
temperature of approximately 23,000 K, we have collisional excitation to the n=3 level, 
but also to n=4. The life-times of the levels n=3,4 are shorter than n=2, so we 
have the production of H$\alpha$ and H$\beta$ line. Even if the life-time of the 
level n=2 can be also short, this level is constantly populated by Ly$\alpha$ 
scattering. Therefore, the population of the energy levels in the Balmer line series
 may be described with the Boltzmann distribution, defined by the excitation temperature.

The line profiles of the Balmer lines in the observed ``BP objects''
are similar (Fig. \ref{sm}), indicating that these lines are close to 
being optically thin. Therefore, we  can state that the objects that follow the BP,
the so-called ``BP objects'' have broad-line hydrogen Balmer lines that
 may originate from the BLR clouds of high density $\sim10^{14} \rm cm^{-3}$, 
in case the broad lines are not influenced by the intrinsic 
dust effects. Moreover, these so-called ``BP objects'' tend to have broader 
BEL indicating that the regions where the BP fit works
should have higher emitter-velocities and be closer to the central 
engine \cite[see Fig. 9 in][]{LM07}.

Finally, it must be noted that the BLR plasma cannot be generally 
considered to be in the equilibrium regime for which the Boltzmann distribution
can be used, since the density of emitters is probably not uniform 
across the region. The density can be expressed as \citep{OF06}:
$N_m \sim b_m(T, n_{\rm e}) g_m \exp(-E_{mn}/kT)$, 
where $b_m(T, n_{\rm e})$ represents the deviation from thermodynamical 
equilibrium. In general, photoionization models predict that 
$b(T, n_{\rm e}) \neq 1$, but, taking into account the ratio of emission 
lines belonging to the same series, the deviation from thermodynamical 
equilibrium might be overtaken if the condition 
$b_i(T, n_{\rm e}) / b_j(T, n_{\rm e}) \approx 1$ is met, which
should be possible at least for higher energy levels \citep{OF06}.

\section{Conclusions}

In this paper we present a study of the Balmer line ratios
of AGN and their usage as a tool to investigate physical properties of the 
broad line region in AGN. The main results are based on the analysis 
of the broad hydrogen Balmer lines (H$\alpha$ to H$\varepsilon$).
The lines are obtained from the photoionization models generated by 
the spectral synthesis code CLOUDY (version C08.00), calculated  using 
the recombination theory for hydrogenic ions, and measured from the sample 
of observed spectra of 90 broad-line AGN from SDSS. The Boltzmann-plot 
is applied on all samples and the $T_{\rm BP}$ is obtained. The properties
of spectra with the BP error $f\leq10\%$ and plasma conditions for their formation 
are further explored. From our analysis, we come to these conclusions:

\begin{itemize}

\item[i)] from the CLOUDY photoionization models we obtain that for a limited 
space of physical parameter ($n_{\rm H}$, $\Phi_{\rm H}$) the physical processes 
in plasma are such, that the Balmer lines follow the BP. This is
the case for either very high ionization parameter (low density and high input
ionization flux) or for high densities at higher column densities;

\item[ii)] the recombination theory for hydrogenic ions predicts that the BP 
 may be applicable for higher densities $\sim 10^{14} \rm cm^{-3}$;

\item[iii)] from the analysis of the sample of broad line AGN from the SDSS database, 
it follows that the BP can be applied in $\approx$35\% of cases, and the obtained 
BLR temperatures are in the range $T_{\rm BP}=10,300-25,200$~K.

\end{itemize}

There is a fraction of objects that follow the BP (``BP objects'') indicating
that in their BLR the physical conditions provide that the populations of 
the hydrogen upper energy states ($n \geq 3$) follow the Saha-Boltzmann distribution.
These objects are thus interesting and should be further investigated (e.g. what are
their UV or X-ray spectral characteristics, or what is the connection with
other AGN types) and that will be the subject of our future work.

\begin{acknowledgements}
D.I. would like to thank to the Department of Physics and Astronomy of the
University of Padova, Italy for their hospitality. Also, we thank the 
anonymous referee for useful comments. This work was
supported by the Ministry of Education and Science of the Republic
of Serbia through the project Astrophysical Spectroscopy of
Extragalactic Objects (\#176001). 

Funding for the SDSS and SDSS-II has been provided by the Alfred P.
Sloan Foundation, the Participating Institutions, the National
Science Foundation, the U.S. Department of Energy, the National
Aeronautics and Space Administration, the Japanese Monbukagakusho,
the Max Planck Society, and the Higher Education Funding Council for
England. The SDSS is managed by the Astrophysical Research
Consortium (ARC) for the Participating Institutions. The SDSS Web
Site is http://www.sdss.org/. This research has made use of NASA's
Astrophysics Data System.
\end{acknowledgements}

\end{document}